# Reaction Kinetics in a Tight Spot

Ofer Biham [1], Joachim Krug [2], Azi Lipshtat [1] and Thomas Michely [3]

[1] *Racah Institute of Physics, The Hebrew University, Jerusalem 91904, Israel*
[2] *Institut für Theoretische Physik, Universität zu Köln, 50937 Köln, Germany*
[3] *I. Physikalisches Institut, RWTH Aachen University, 52056 Aachen, Germany*

**Abstract:** The standard analysis of reaction networks based on deterministic rate equations fails in confined geometries, commonly encountered in fields such as astrochemistry, thin film growth and cell biology. In these systems the small reactant population implies anomalous behavior of reaction rates, which can be accounted for only by following the full distribution of reactant numbers.

It is a common experience in society and nature that things become predictable in large populations: The outcome of an election can be forecast by sampling, and the gas pressure in a container can be accurately computed from thermodynamics. This observation, which is a consequence of the law of large numbers, underlies the textbook treatment of reaction rates in chemical and biological systems. To fix ideas, consider some liquid solution containing certain concentrations $\rho_A$ and $\rho_B$ of atomic or molecular species A and B that react with each other according to A + B → C, where C is the reaction product. Each concentration is the *average* number of molecules in a reference volume. According to standard chemical kinetics, C molecules are produced at rate $R = K\rho_A\rho_B$, where the rate constant $K$ is determined by the diffusivity and reactivity of the A and B species. This is the basis of rate equation theory, which applies universally in the case of macroscopic populations.

However, recently researchers in fields as diverse as astrochemistry, surface science and cell biology have independently come to realize that in the confined geometries they encounter, the standard rate equation treatment does not apply. In systems that are partitioned into units of sub-micron size (such as dust grains in an interstellar cloud, islands on a solid surface and cells in a living organism), the average number of reactants in each unit may go down to one or less. In this limit the discrete nature of the reactant populations becomes crucial and the reaction rates are dominated by fluctuations. While it is less remarkable that, under these circumstances, the reaction becomes a noisy process, the unexpected finding is that also the *average* reaction rate deviates strongly from the standard prediction, because the confinement introduces strong correlations between the reactants. This observation was made in two areas that are, literally, light years apart: Interstellar chemistry, and thin film deposition by molecular beam epitaxy. In this article we would like to argue that similar effects should be relevant to a broad range of nanoscale systems. After describing the two key examples, we conclude with some specific remarks aimed at reaction networks in biological cells.

**Interstellar Chemistry.** Since hydrogen is the most abundant chemical element in the Universe it is not surprising that $H_2$ is the most abundant molecule. But molecular hydrogen formation in gas-phase reactions is highly inefficient. In interstellar clouds, the dominant production mechanism of $H_2$ molecules takes place on the surfaces of dust grains. In this process, hydrogen atoms collide and stick to grain surfaces, on which they diffuse. When two atoms encounter each other they form a molecule. Given the low density of interstellar clouds and the sub-micron size of the grains, on each grain there may be at most a handful of atoms.[1] Under these conditions the rate equations fail and the *master equation approach*, which accounts correctly for the reaction rate even when fluctuations are dominant, is called for.

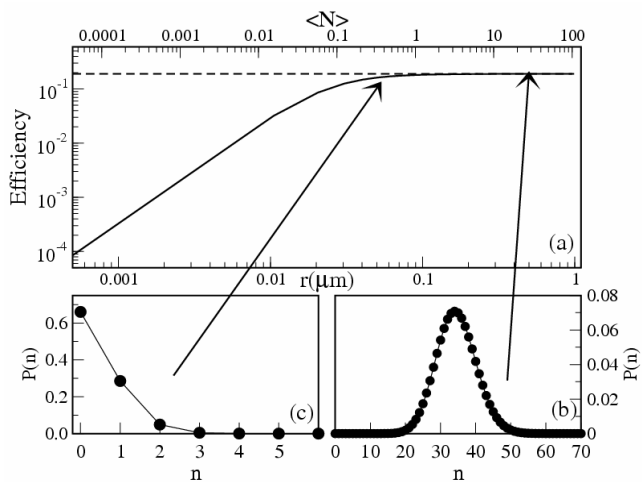

**Figure 1:** (a) The production efficiency of $H_2$ molecules (the fraction of H atoms adsorbed on the grain that end up in molecules), obtained from the master equation (solid line) and from the rate equation (dashed line) vs. the grain radius (assuming spherical grains) for olivine grains at 10 K. [3] Below a radius of about 0.1 $\mu m$, the average number $\langle n \rangle$ of atoms on a grain (upper scale) becomes smaller than 1 and the production efficiency starts to deviate from the rate equation results. The distribution $P(n)$ of the number of atoms on a grain is shown for a large grain (b) and a small grain (c).

The master equation keeps track of the probabilities $P(n)$, $n = 0,1,2,\ldots$, to have $n$ hydrogen atoms simultaneously on a grain.[2,3] For a single grain, the reaction rate is proportional to the number of pairs of atoms, namely to $n(n-1)$. In particular, the reaction rate is zero when $n = 1$ because it takes two atoms to form a molecule. Averaging over an ensemble of grains, $R \sim \langle n(n-1) \rangle$. In contrast, the rate equations assume the reaction rate to be proportional to the square of the concentration of atoms, i.e. $R \sim \langle n \rangle^2$, where $\langle n \rangle$ is the average number of atoms on a grain. If the grains were connected as patches of a macroscopic surface, $P(n)$ would be a Poisson distribution and the two expressions would be identical, because then $\langle n^2 \rangle - \langle n \rangle^2 = \langle n \rangle$. In contrast, on the surface of a small grain, where the probability for two reactants to meet before one of them desorbs is close to unity, the reaction itself strongly depletes the probability to find *more* than two atoms simultaneously. Then $P(n)$ becomes distinctly non-Poissonian and $\langle n(n-1) \rangle \ll \langle n \rangle^2$ which implies that rate equation theory greatly overestimates the reaction rate (Fig. 1).

**Thin Film Growth.** The formation of thin films by molecular beam epitaxy takes place through diffusion of the deposited atoms on the surface, followed by island nucleation and growth.[4] The nucleation process resembles the formation of molecules, since it takes two atoms (or more) that encounter each other to form a stable cluster. After the growth of the first layer, nucleation typically takes place on top of islands, which are bounded by step edges. Repeated nucleation on top of freshly formed islands leads to the formation of pyramidal mounds (Fig. 2). Due to an energy barrier which prevents atoms from descending across the step,[5] the atoms on the top terrace are confined long enough to repeatedly visit each site on the terrace. The problem then becomes essentially equivalent to molecular hydrogen formation on small grains.[6]

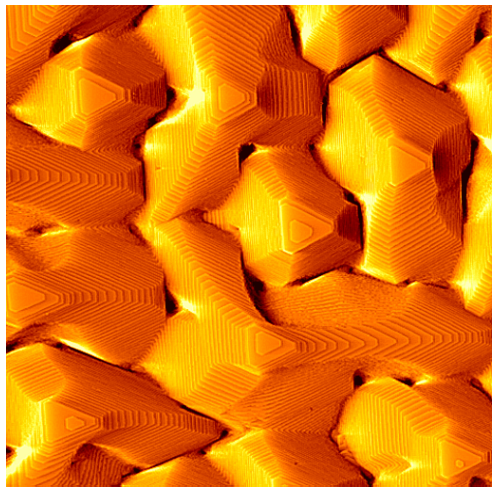

**Figure 2:** Scanning tunneling microscope image (345 × 345 nm) of the surface of a crystalline platinum film after deposition of 93 atomic layers at a temperature of 440 K. New islands nucleate on top of the pyramidal mounds and grow until they reach the size of the underlying triangular terrace. Top terraces of different sizes represent different stages in this growth process.

The rate at which new islands nucleate is directly observable in images such as Fig.2, because it determines the size of the top terrace of a mound.[7] In the situation shown in Fig. 2, the average number of atoms on a top terrace is only a few times $10^{-4}$. This is far below the range of validity of rate equation theory, which would therefore grossly overestimate the nucleation rate. In response to this insight, a refined statistical theory of the nucleation process has been developed,[7-9] which provides a detailed description of the shape of the growing mounds in terms of atomistic parameters.[4]

**Reaction Networks in Cells.** Chemical reactions involving small numbers of molecules abound in cell biology, and the resulting stochastic fluctuations have been a concern in the field for a long time.[10-12] Constituents such as mRNAs or even proteins strongly fluctuate in number, and some of them have averages of order 1 or less per cell. Fluctuations in the copy numbers of certain mRNA molecules affect the production rates of the resulting proteins. These fluctuations may be further amplified through specific genetic pathways, and they are relevant to many basic cellular processes such as cell division, and to the emergence of heterogeneity in cell populations. Model calculations for autocatalytic reaction cycles have shown that the presence of species with a very small number of molecules can induce *qualitatively* novel states which do not occur in a rate equation treatment.[13]

The *quantitative* characterization of noise in biological systems requires sophisticated measurements that are now becoming possible.[14] This has inspired much recent work on fluctuations both in the temporal behavior of a single cell and across cell populations, which utilizes chemical master equation techniques.[15,16] Because all reactions take place in the confinement of the cell membrane, the mechanism described above could result in anomalous distributions, as well as in deviations of the averages from the predictions of macroscopic models.

Further progress will require to face the experimental challenge of finding the structure of the reaction networks and evaluating the large number of rate constants, as well as the theoretical challenge of solving complex master equation systems with thousands of coupled equations. A promising methodology for the simulation of such systems was recently introduced in the context of gas-grain chemistry.[17] This illustrates well how the cross fertilization between astrochemistry, surface science and cell biology is already contributing to the evolution of this emergent, interdisciplinary field of research.

**Acknowledgements:** We thank Nathalie Q. Balaban for helpful discussions. Support by the Israel Science Foundation, the Adler Foundation for Space Research, and Deutsche Forschungsgemeinschaft within SFB 616 *Energy Dissipation at Surfaces* is gratefully acknowledged.